\documentclass[%
 reprint,
 amsmath,amssymb,
 aps,
]{revtex4-2}

\usepackage{url}
\usepackage[%
  colorlinks=true,
  urlcolor=blue,
  linkcolor=blue,
  citecolor=blue
]{hyperref}

\usepackage{placeins} 
\usepackage{xcolor}

\usepackage{xparse}

\NewDocumentCommand{\tens}{t_}
 {%
  \IfBooleanTF{#1}
   {\tensop}
   {\otimes}%
 }
\NewDocumentCommand{\Log}{o}{%
  \IfNoValueTF{#1}{}{{}^{#1}\!}\log}%

\newcommand{\ket}[1]{| #1 \rangle}

\usepackage{graphicx}
\usepackage{dcolumn}
\usepackage{bm}

\begin{document}

\title{Genuine Parrondo's paradox in quantum walks with time-dependent coin operators}
 
\author{Marcelo A. Pires$^{1}$}
\thanks{piresma@cbpf.br}

\author{S\'{\i}lvio M. \surname{Duarte~Queir\'{o}s}$^{1,2}$}
\thanks{sdqueiro@cbpf.br}

\affiliation{
$^{1}$Centro Brasileiro de Pesquisas F\'isicas, Rio de Janeiro/RJ, Brazil
\\
$^{2}$National Institute of Science and Technology for Complex Systems, Brazil}

\date{\today}

\begin{abstract} 
We show that a genuine Parrondo paradox can emerge in two-state quantum walks without resorting to experimentally intricate high-dimensional coins. To achieve such goal we employ a time-dependent coin operator  without breaking the translation spatial invariance of the system.
\end{abstract}

\maketitle

\section{\label{sec:intro}Introduction}

 From the so-called Brazil nut effect in granular materials~\cite{brazilnuteffect} -- where the largest particles of variously sized blend end up on its surface when subjected to (random) shaking -- to the boosting of the long-term growth rate of a population by allocating offspring to sink habitats~\cite{jansen}, Nature has provided us with a myriad of instances which defy common sense and thus are often understood as paradoxical~\cite{paradoxes}. Within this class of systems yielding counterintuitive results, we can also refer to several thermodynamical approaches that attempted to come up with perpetual machines of both first and second kind. A canonical instance thereof is the well-known Feynman's ratchet and pawl machine~\cite{feynman-lecphys}
(later scrutinized in Ref.~\cite{parrondo-critique}).
The concept of ratchet was later employed to Brownian particles in a periodic and asymmetric potential that systematically moves to one of the sides when potential is switched on and off~\cite{hanggi-parrondo}.
Such a mechanism was later reinterpreted from a gambling perspective paving the way to the assertion that the combination of two losing games can yield a winning game when combined. That understanding was later honed to a scenario related to good and bad biased coins which are played more or less frequently when the two games are combined. That recast of a winning combination of losing games case was coined Parrondo's paradox~\cite{Parrondo1996,HarmerAbbott1999,harmer2002review,Abbott2010,parrondo2004brownian}. The so-called Parrondian phenomena have lured the information theory community -- at first, for its connection with random number generation and game theory -- and reached the quantum realm for the development of quantum ratchets, walks (QW) and quantum games that can be translated into a Parrondo framework. As in classical systems, there are different variants of the quantum Parrondo effect~\cite{lee2002quantum,chen2010quantum,zhu2011quantum,li2011quantum,wang2011parity,lai2020parrondo,banerjee2013enhancement,strelchuk2013parrondo} (for a recent review see Ref.~\cite{laiparrondo}).
In Ref.~\cite{toral2001cooperative}, Toral introduced an alternative classical Parrondo walk, the cooperative Parrondo's games, that subsequently gained quantum 
versions~\cite{bulger2008position,pawela2013cooperative,miszczak2014general}. 

In this manuscript, we aim at implementing an actual Parrondo strategy with Quantum Walks(QWs)~\cite{aharonov1993quantum} as they are multi-purpose models with several possibilities for experimental realizations~\cite{wang2013physical,Gr_fe_2016,flamini2018photonic,neves2018photonic} and links with both fundamental~\cite{wu2019topological} and 
applied~\cite{ambainis2003quantum,portugal2013quantum,venegas2008quantum,janexperimental} studies.
To the best of our knowledge, the first attempts 
to set forth a Parrondian QW -- considering a capital-dependent rule implemented with a position-dependent potential -- were conveyed in Refs.~\cite{MeyerBlumer2002_JSP,MeyerBlumer2002_FNL,meyer2003noisy}. In spite of being successful in the short-run, a long-run analysis shows their Parrondo's paradox can be temporarily suppressed due to periodicity in expected payoffs.
Other close attempts at implementing Parrondo's  paradox with QWs failed in the asymptotic limit~\cite{Flitney2012,li2013quantum} as well. Taking a rather different road, it was shown in Ref.~\cite{ChandrashekarBanerjee2011} the emergence of a Parrondo-like effect consisting of the obtention of an unbiased game from alternating biased games using QWs. However, the issue of a QW-based implementation of a genuine Parrondo game remained pending irrespective of some proposals~\cite{flitney2004quantum,gawron2005quantum,Kosik2007,Rajendran2018_EPL,rajendran2018implementing,Machida_Grunbaum2018} that demand high-dimensional QWs, which are harder to implement than the qubit-based instances. For example, in Ref.~\cite{flitney2004quantum}, it was used a multi-coin approach with history-dependence and in Ref.~\cite{gawron2005quantum,Kosik2007} the authors opted for a multi-register protocol.  More recently, it was used a three-state QW~\cite{Rajendran2018_EPL} and in Ref.~\cite{rajendran2018implementing,Machida_Grunbaum2018} it was used a two-coin QW.

That said, we have verified the implementation of a genuine Parrondo paradox within the scope of 2-state QWs with simple alternations between single-parameter coins remains open so far. To solve it, we have resorted to a time-dependent coin operator, which exhibits a very rich phenomenology~\cite{banuls2006quantum,vieira2013dynamically,di2015massless} alongside time-dependency on the translation operator~\cite{di2018elephant,pires2019multiple,pires2019quantum}. Hereinafter, we assert that in breaking the temporal constancy of the coin operator in the QW it is possible to successfully implement a quantum Parrondo's paradox.

\begin{figure*}[t]
  \centering

\includegraphics[width=0.99\linewidth]{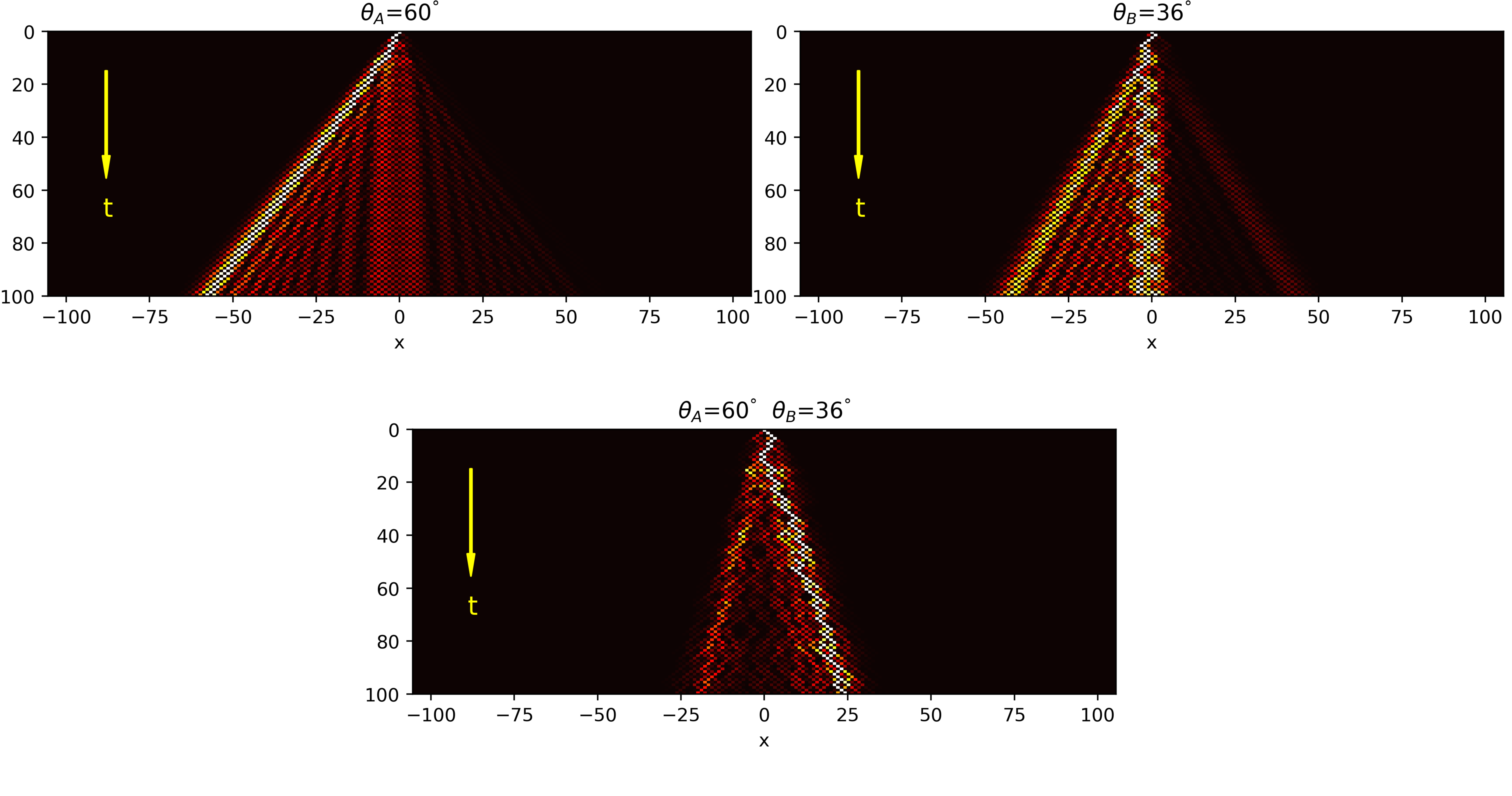}

\caption{Spatiotemporal evolution of the normalized $P_t(x)/P_t^{\max}$ for $t=100$. Genuine Parrondo's  paradox with the generalized hadamard coin with the novel protocol  $\theta_t=(t+1)\theta$ for a 2-state quantum walk: a combination of losing strategies (left-biased $P_t(x)$) becomes a winning strategy(right-biased $P_t(x)$). Different from previous works, this suggests it is not necessary to employ high-dimensional quantum states to implement a Parrondo's  game. Up-left: $\theta=60^o$. Up-right: $\theta=36^o$. Down: switching $\theta=60^o$ and $\theta=36^o$.}
\label{fig:pxt2d}
\end{figure*}

\section{\label{sec:model}Model}

\bigskip

Explicitly, our model goes as follows: at a given time $t \in \mathbb{N}$, we consider a QW with a full wave function given by $\Psi_t$ as
\begin{align}
\Psi_t 
=
\sum_{x \in \mathbb{Z}} 
\left(
\psi_t^U(x) 
\ket{U}
+
\psi_t^D(x) 
\ket{D}
\right)
\otimes
\ket{x} 
\label{Eq:psi_geral}
\end{align}
where $\{U,D\}$ (standing for up and down, respectively) is the internal degree of freedom of our two-state quantum walker moving in $x \in \mathbb{Z}$, which corresponds to its external degree of freedom. That is to say, our QW lives in the composite Hilbert space  $\mathcal{H}_2\otimes \mathcal{H}_\mathbb{Z}$. The functions $\psi_t^{U,D}(x)$ are the spatio-temporal amplitude of probability associated to $\{U,D\}$, respectively. The evolution $t \rightarrow t+1 $ proceeds with the application of the operator $\hat W$ as 
\begin{align}
\Psi_t  \xrightarrow{ \hat W_t} \Psi_{t+1}
\label{Eq:WSC1}
\end{align}
\begin{align}
\hat W_t = \hat{T}(\hat{R_t}\otimes \mathcal{I}_\mathbb{Z})
\label{Eq:WSC2}
\end{align}

with the identity operator $\mathcal{I}_\mathbb{Z}=\sum_{x \in \mathbb{Z}} |x \rangle \langle x|$ and: 
\begin{itemize}
    \item The coin operator:
  \begin{align}
    \begin{cases}
     |x, U \rangle
     \xrightarrow{\widehat{R}}
     c_{UU}(t) |x, U \rangle +
     c_{DU}(t) |x, D \rangle
      \\
     |x, D \rangle 
     \xrightarrow{\widehat{R}}
     c_{UD}(t) |x, U \rangle +
     c_{DD}(t) |x, D \rangle
    \end{cases}
    \label{Eq:C-explicito}
  \end{align}
   where $c_{ij}$ that are the elements of a rotation matrix  that will be described shortly. 
   
    \item  The state-dependent shift operator: 
  \begin{align}
    \begin{cases}
     |x, U \rangle 
     \xrightarrow{\widehat{T}}
     |x+1, U \rangle 
      \\
     |x, D \rangle 
     \xrightarrow{\widehat{T}}
     |x-1, D \rangle 
    \end{cases}
    \label{Eq:T-explicito}
  \end{align}

\end{itemize}

For the coin operator, we choose a generalized version of the Hadamard operator
\begin{equation}
\widehat C_{H}(t)
= 
\cos \theta_t \widehat\sigma_z + \sin\theta_t \widehat\sigma_x
\end{equation}
where $\widehat\sigma_{z}$ and $\widehat\sigma_{x}$ are the standard Pauli matrices.
Based on Ref.~\cite{panahiyan2018controlling}, we choose $\theta_t$ as a linear function of time, 
namely $\theta_t = (t+1)\theta$, the linearity of which has the advantage of being feasible for experiments. 

Following the literature related to the Parrondian phenomenon~\cite{parrondo2004brownian,Abbott2010,Flitney2003,laiparrondo} we have assumed the switching rule: for $t$ even we apply $\theta_A$ otherwise we applied $\theta_B$.
Concerning $x$, there are two possible interpretations in this manuscript that we have used interchangeably. On the one hand, when $x$ was set as capital~\cite{MeyerBlumer2002_JSP,MeyerBlumer2002_FNL}, then the QW was interpreted as a game where $x>0$ means a positive payoff. 
Taking Refs.~\cite{boman2001parrondo,piotrowski2004quantum} into account, such interpretation of a capital-dependent game may provide new insights within the context of quantum-like modeling of financial processes~\cite{piotrowski2017quantum}. This perspective gives further importance to the development of novel quantum games, as we intend to carry out here. On the other hand, setting $x$ as the position of a particle we shed light on transport phenomena by we worked at computing the probability current towards $x<0$ or $x>0$.

\begin{figure*}[t]
\centering
\includegraphics[scale=0.57]{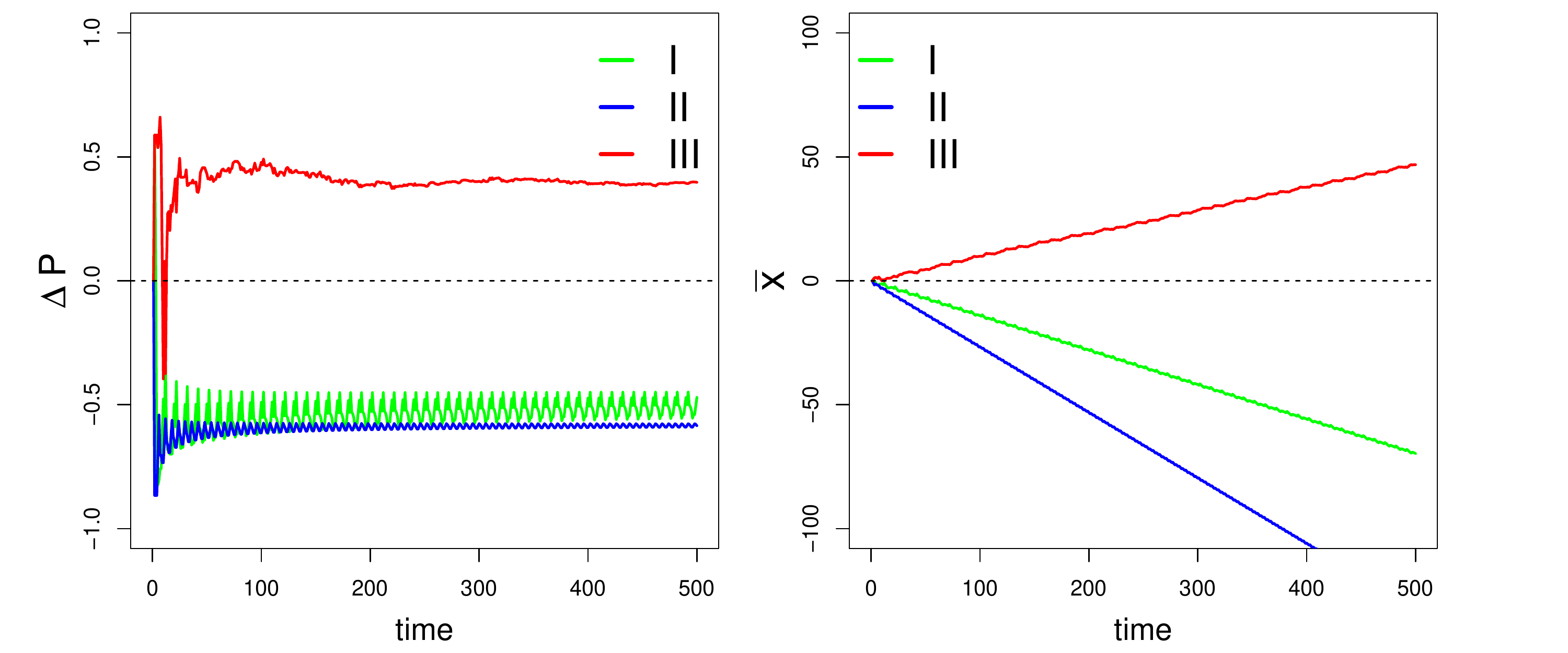}
\caption{Time series for the expected payoff $\Delta P= P_L-P_R$ and  for the mean position $\overline{x}(t) = m_1(t) $. The net directed current becomes reverted in the opposite direction under the alternating prescription. I: only $\theta=60^o$. II: only $\theta=36^o$. III: alternating $\theta=60^o$ and $\theta=36^o$. }
\label{fig:pr_pl_and_xmean}
\end{figure*}

\begin{figure}[t]
  \centering
\includegraphics[scale=0.47]{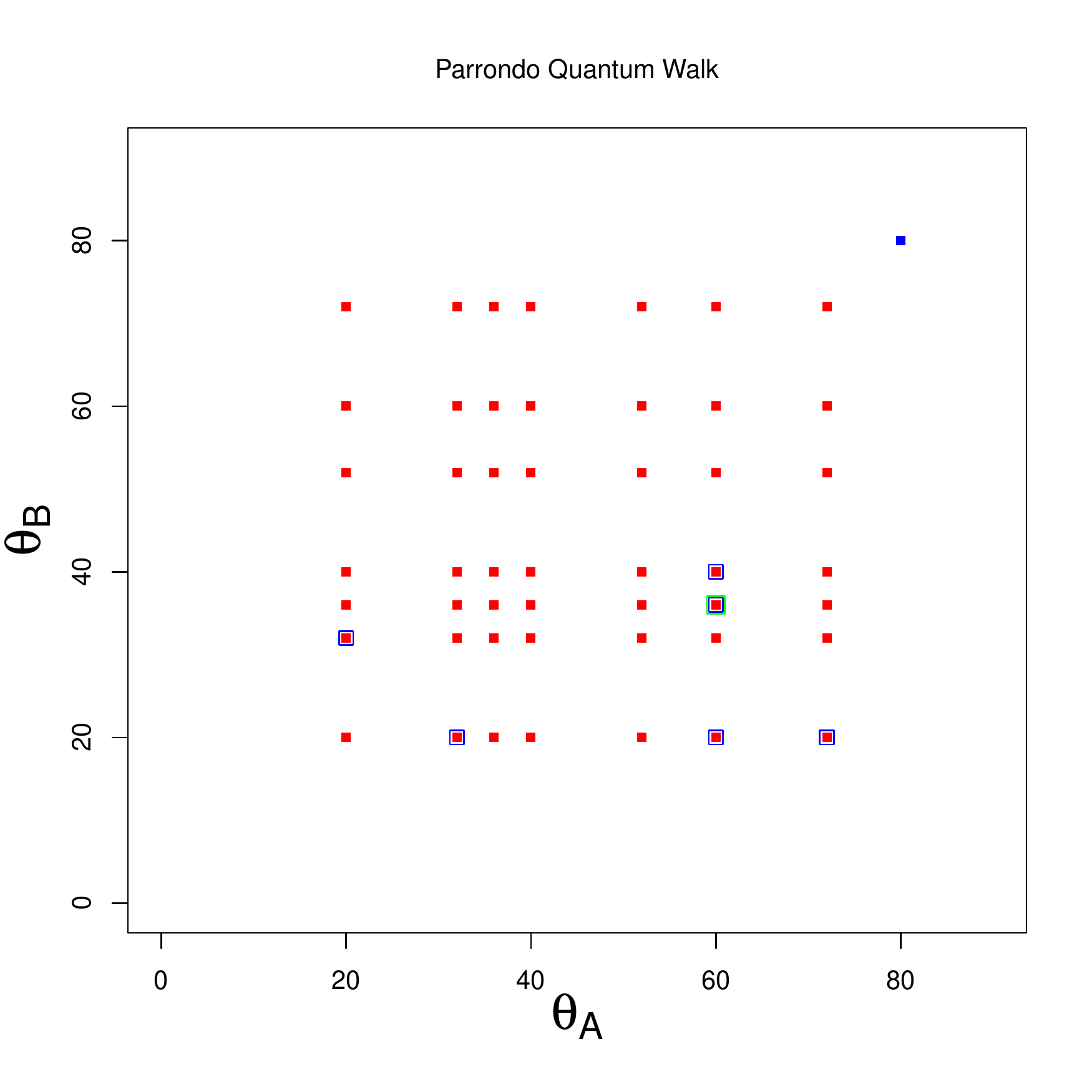}  
\caption{Diagram for the Parrondo QW. The red squares indicate the left-skewed scenarios with $\Delta P^{\theta_A}< -\epsilon$  and  $\Delta P^{\theta_B}< -\epsilon$. The  blue square indicate right-skewed scenarios with $\Delta P^{\rm combined}>\epsilon$. We set $\epsilon=1/3$ in order to avoid artifacts. The current reversal takes place in the cases with red and blue squares. The green open square indicates the case corresponding to the Figs.~\ref{fig:pxt2d}-\ref{fig:pr_pl_and_xmean}. Simulations performed with $t_{\max}=1000$. All quantities in this diagram are temporal averages discarding the $t_{\rm transient}=t_{\max}/2$.}
\label{fig:diagram}
\end{figure}

\section{\label{sec:results}Results and discussion}


To analyze the outcome of this dynamics we have first considered the probability distribution 
\begin{equation}
P_t(x) = |\psi_{t}^{D}(x)|^2 + |\psi_{t}^{U}(x)|^2,
\end{equation}
whence we computed
\begin{equation}
P_t^{\max}={\max}_{x} P_t(x)
\end{equation}
which is the maximum of $P_t(x)$ over the chain at a time $t$. 
In Fig.~\ref{fig:pxt2d}, we show how $P_t(x)/P_t^{\max}$ evolves over time. 
For $\theta_A$, we see a left-biased flux of probability. 
For $\theta_B$, we observe a left-biased current of probability as well.
Interestingly, the alternation between  $\theta_A$ and  $\theta_B$
leads to a  counterintuitive phenomenon corresponding to the very nature of Parrondo's paradox: the combination of two losing games ($P_t(x)$ towards $x<0$) gives rise to a winning strategy ($P_t(x)$ towards $x>0$). 
In other words, in such quantum carpets, the coupling of two protocols with a left-biased current of probability ends up producing a right-biased current. 

In the left panel of Fig.~\ref{fig:pr_pl_and_xmean}, we quantify the current of probability
\begin{equation}
\Delta P =  P_L-P_R=\sum_{x<0}P_t(x)-\sum_{x>0}P_t(x)
\end{equation}
where in the combined protocol (III) we see a negative peak for $t<20$. Afterward, the current of the combined game becomes robustly positive.

In the right panel of Fig.~\ref{fig:pr_pl_and_xmean}, we plot the first ($n=1$)-order statistical moment
\begin{equation}
m_n(t) = \overline{x^n}(t) =\sum_x x^nP_t(x),
\end{equation}
from which it is possible to perceive the emergence of the Parrondo's effect.

At this point, we still have to answer the question regarding the underlying mechanism yielding the quantum Parrondo effect. First, let us recall the canonical reasoning explaining the aforementioned paradox:  blending a game composed of advantageous and disadvantageous ingredients with a losing game, inhibits the unfavorable part in the first game. In the standard QW, the central component is quickly suppressed because its probability, $P_t(x)$, concentrates in the borders. Yet, in our protocol, the central component of $P_t(x)$ remains for each isolated game, for instance, with $\theta=60^o$ and $\theta=36^o$. This is similar to the `good' (favorable)   and `bad'(unfavorable) components. Therefore, we learn the alternating protocol ($\theta=60^o$ and $\theta=36^o$) hinders the left-biased borders, which prompts the prevalence of the right component, as shown in Fig.~\ref{fig:pxt2d}.

Finally, in Fig.~\ref{fig:diagram} we see a diagram with $\theta_A$ versus $\theta_B$ for runs until $t_{\max}=10^3$ where we discard the transient $t<t_{\max}/2$ to compute the temporal average of current of probability $\Delta P^{\rm single}$ for the single game (with solely $\theta_A$ or $\theta_B$) as well as for 
$\Delta P^{\rm combined}$ relatively to the combined game (alternating $\theta_A$ and $\theta_B$). It is visible there are specific combinations that lead to the Parrondian effect. From that diagram, we understand that we do not have a smooth Parrondo line, but a set of points wherein a Parrondo strategy emerges. This occurs in other quantum Parrondo-like cases as well and we ascribe it to the complexity of the quantum system we are treating. Contrarily to classical systems, the state of a quantum particle is ruled by the superposition principle, and thus only very specific combinations of the two coins, i.e., $\theta_A$ and $\theta_B$ allow obtaining an overall winning strategy.

\section{\label{sec:remarks}Concluding remarks}


In conclusion, one should acknowledge the Parrondo's paradox contains an important lesson to be learnt: one must be careful before labelling a given protocol as useless, because from that it is possible to create a combined protocol -- as we have made here -- that ends up yielding the features one is aiming at. Apart from that, taking into consideration the assertion in Ref.~\cite{rajendran2018implementing} the need for investigating the class of Parrondian phenomena via QWs is prompted by the research to improve quantum algorithms, namely search algorithms.

The classical Parrondian paradigm can be introduced within the scope of game theory. Similarly the quantum Parrondo's  paradox can be introduced in the realm of the Quantum Game Theory~\cite{Klarreich2001,LeeJohnson2002,guo2008survey,khan2018quantum,huang2018survey}. 
In this sense and taking into account that QWs are versatile quantum simulators~\cite{aspuru2012photonic}, our 
new protocol for implementing a Parrondo's  game can be employed as a platform to provide further insights in the Quantum Game Theory both theoretically and experimentally. 

One advantage of our protocol to achieve directed transport is that we do not require a breaking in the spatial symmetry associated to the coin operator. This approach contrasts with the state-of-the-art like Ref.~\cite{chakraborty2017quantum} where 
spatial invariance is broken by introducing a pawl-like effect with position-dependent coin operations. That is, their walk operator is embedded with a local spatial asymmetry. Another advantage of  this protocol is the straightforward use of qubits, a feature that makes the model feasible for implementation in photonic architectures~\cite{wang2013physical,Gr_fe_2016,flamini2018photonic,neves2018photonic}. More specifically the setups introduced in Ref.~\cite{xue2015experimental} are quite appropriate to accommodate this time-dependent coin operator with proper adjustments. From an experimental perspective, our protocol contributes with  a fresh prospect for laboratorial implementation of Parrondo’s games in a physical systems beyond the scope of the original model~\cite{si2012optical} as it fills a gap in that domain by successfully obtaining the actual Parrondo effect with 2-state qubit Parrondian model in employing a time-dependent coin. Still, the findings we have reported suit the development of new devices for current reversal without the application of an external gradient. From the point of view of QWs, our work add a novel application of time-dependent coin operators.

\smallskip
\begin{acknowledgments}
We acknowledge  financial support from the Brazilian funding agencies CAPES (MAP) as well as CNPq and FAPERJ (SMDQ).
\textcolor{black}{Whilst we were writing up this manuscript, the preprint Ref.~\cite{jan2020study} came out where a Parrondo QW  with  intricate alternations between three-parameter coins is discussed.}
\end{acknowledgments}

\FloatBarrier
\bibliography{apssamp.bib}


\end{document}